\documentclass[12pt,a4paper]{article}

\usepackage{amssymb}
\usepackage{amsmath}
\usepackage[version=4]{mhchem}
\usepackage{hyperref}
\usepackage{pdflscape}
\usepackage{makecell}
\usepackage{longtable}
\usepackage{adjustbox}

\begin{document}

\title{Magnetic Properties of the Quasi-1D Magnesium Lanthanide Borates \ce{Mg\textit{Ln}B5O10}}

\author{Lachlan G. M. Rooney, Si\^{a}n E. Dutton, Nicola D. Kelly\thanks{ne281@cam.ac.uk}}

\date{Cavendish Laboratory, University of Cambridge,\\ J J Thomson Avenue, Cambridge, CB3 0US, UK}

\maketitle

\begin{abstract}
Lanthanide borates are widely studied for their optical and magnetic properties. A wide variety of structures are known with 3, 2, 1 and 0 dimensional connectivity of lanthanide ions. Here, we explore \ce{Mg\textit{Ln}B5O10}, with a quasi-1D arrangement of the \textit{Ln} ions. Polycrystalline samples of \ce{Mg\textit{Ln}B5O10} (\textit{Ln} = La, Pr, Nd, Sm--Er) were synthesised with high purity via a sol-gel method. Powder X-ray diffraction data confirmed the reported monoclinic space group ($P2_1/c$). The magnetic $Ln^{3+}$ ions in \ce{Mg\textit{Ln}B5O10} form relatively isolated zig-zag chains parallel to the $b$ axis. Magnetic susceptibility and isothermal magnetisation were measured: all samples except $Ln=$ (Eu, Sm) fit the Curie-Weiss Law in isothermal magnetisation at high temperatures, in broad agreement with theoretical expectations. $Ln =$ (Nd, Tb, Dy, Ho) exhibit signatures characteristic of Ising spin saturation, implying single ion anisotropy, while Gd exhibits characteristics of Heisenberg spins. Estimation of magnetic interactions suggests that \ce{Mg\textit{Ln}B5O10} are candidate materials for quasi-1D magnetism. The magnetocaloric entropy change was also calculated, with \ce{MgGdB5O10} showing promise for application to solid-state refrigeration at liquid helium temperatures.
\end{abstract}

\section{Introduction} \label{Intro}

One-dimensional (1D) magnetic systems -- spin chains of magnetic ions -- have been predicted to exhibit a range of exotic magnetic behaviour, for example spinon excitations. \cite{blundell_magnetism_2001} While truly 1D systems cannot be realised in the solid state, quasi-1D behaviour may be observable given sufficient separation between neighbouring magnetic chains. In such systems, weak interactions across chains cause 3D ordering at low temperatures. 

Low-dimensional magnetism has been investigated in a range of systems, including transition metal structures such as $ABX_3$ (where $A$ is a non-magnetic, singly charged cation, $B$ a magnetic, doubly charged cation, and $X$ a halide ion). \cite{blundell_magnetism_2001,kakurai_magnetic_1992,mukherjee_magnetic_2017} Unlike transition metal ions, in lanthanides the superexchange and dipolar (through space) interactions are similar in magnitude, so it is harder to realise quasi-1D magnetic structures. \cite{orchard_magnetochemistry_2003} However, lattice geometry can prevent the simultaneous satisfaction of all magnetic interactions in frustrated systems, inhibiting long-range order and promoting quasi-low-dimensional magnetism. 

For example, the monoclinic lanthanide metaborates \ce{\textit{Ln}(BO2)3} contain chains of $Ln^{3+}$ ions, resulting in frustration and quasi-1D magnetism for some \textit{Ln}. \cite{mukherjee_magnetic_2017} Quasi-1D ferromagnetic ordering has also been observed in \ce{Tb(HCO2)3}, where \ce{TbO9} polyhedra form chains along the $c$ axis. These chains then form a triangular lattice mediated by the formate ligand. Whilst the magnetisation direction alternates between chains, this alternation is frustrated, inhibiting 3D ordering. \cite{harcombe_one-dimensional_2016,saines_searching_2015} 

Meanwhile, 2D ordering has been observed in the lanthanide orthoborates \ce{\textit{Ln}BO3}, where the lanthanide ions lie on a distorted triangular lattice. \cite{mukherjee_magnetic_2018} Another family of 2D lanthanide borates, \ce{\textit{A}Ba\textit{Ln}(BO3)2} (for \textit{A} = Na, K, Rb), has also been widely explored as a model for the undistorted frustrated triangular lattice, although some compositions suffer from non-magnetic site disorder which may affect their magnetic properties. \cite{Guo2019c,Guo2019b,Sanders2017,Guo2019a,Guo2019}

\begin{figure}[htbp]
    \centering
    \includegraphics[width=1.0\linewidth]{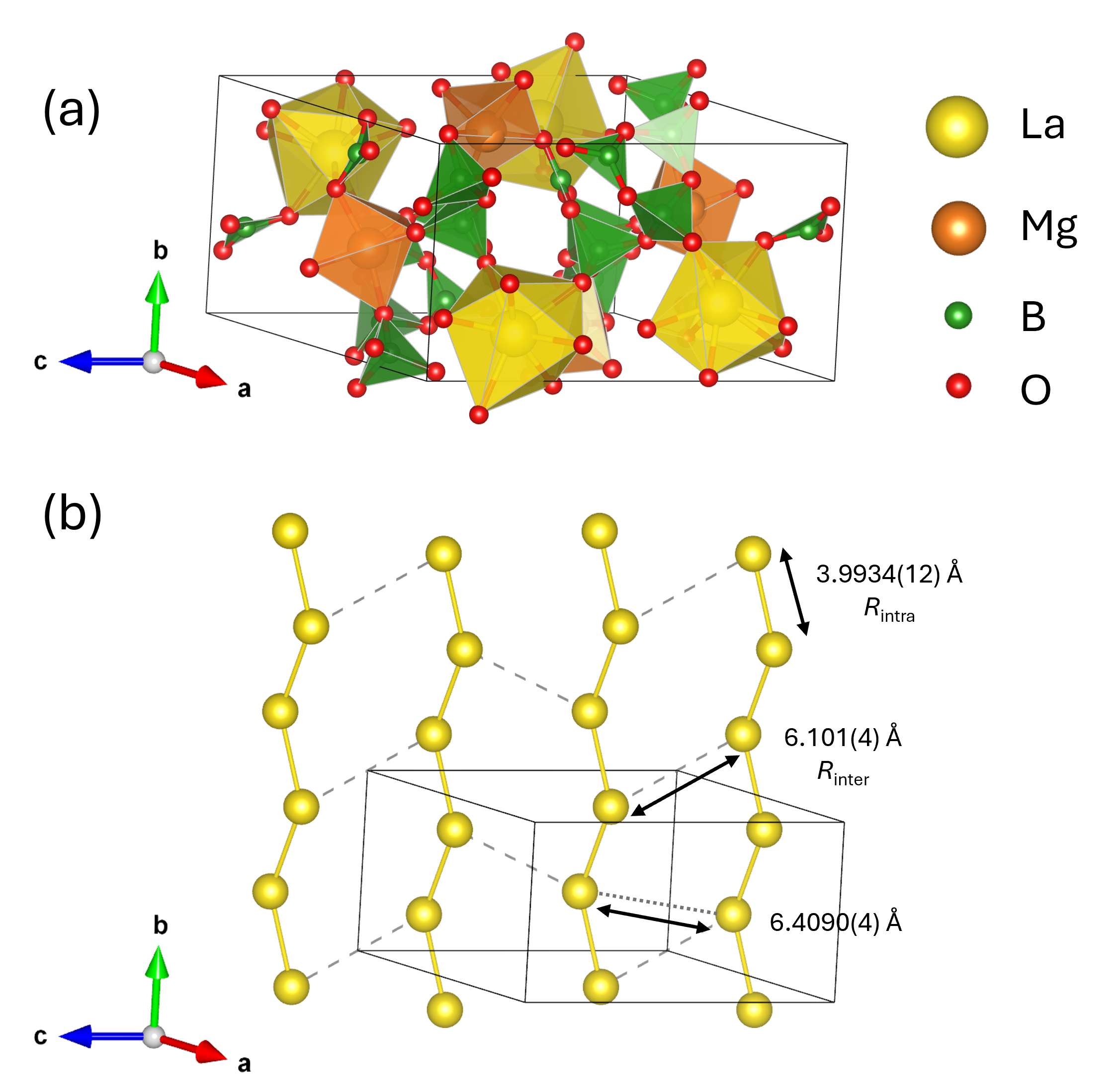}
    \caption{Crystal structure of \ce{MgLaB5O10} in space group $P2_1/c$: (a) all atoms; (b) only the lanthanum ions. The La$^{3+}$ ions form zig-zag chains in the $b$ direction and the distances $R_\mathrm{intra}$ and $R_\mathrm{inter}$ are indicated. Image produced using VESTA. \cite{momma_vesta_2011}}
    \label{fig:unit_cell}
\end{figure}

The magnesium lanthanide borates \ce{Mg\textit{Ln}B5O10} present another opportunity to study low-dimensional magnetism because the magnetic $Ln^{3+}$ ions form relatively isolated chains within the crystal. \ce{Mg\textit{Ln}B5O10} crystallises in the monoclinic $P2_1/c$ space group for \textit{Ln} = (La--Er), \cite{zhang_phase_2017}, figure~\ref{fig:unit_cell}. The \ce{B5O10} form two-dimensional layers linked by the metal ions, with $Ln$ coordination polyhedra forming extended zigzagging chains extended along the $b$ axis. \cite{fouassier_self-quenching_1981,saubat_synthesis_1980} For the La compound, the nearest-neighbour (within a chain) La--La spacing was reported as 3.994 \r{A}, while that across chains was 6.101 and 6.430 \r{A}, largely confining magnetic interactions within the chains. \cite{fouassier_self-quenching_1981,saubat_synthesis_1980} 

The concentration quenching of luminescent emission in the \ce{\textit{Ln}_xLa_{1-x}MgB_5O_{10}} system (\textit{Ln} = Eu, Tb; $0\leq x\leq 1$) was investigated by Fouassier \textit{et al.}, \cite{fouassier_self-quenching_1981} demonstrating the dominance of interactions within (rather than between) $Ln$ chains. As described above, this is expected to produce interesting low-dimensional magnetic behaviour, but the only magnetic study to date was on \ce{MgNdB5O10} which was found to be paramagnetic down to 1.7~K. \cite{cascales_paramagnetic_1999}

The suppression of the 3D ordering temperature in low-dimensional systems motivates their application to solid-state magnetic refrigeration via the magnetocaloric effect (MCE), as an alternative to liquid helium. When the system is adiabatically demagnetised, the magnetic entropy increases as the moments dealign with the field direction, and the corresponding decrease in the lattice and electronic entropy lowers the material's temperature. The long-range magnetic ordering temperature sets a lower cooling limit, so suppressed 3D ordering is needed to access liquid helium temperatures. \cite{balli_advanced_2017,romero_gomez_magnetocaloric_2013}

The current benchmark material for magnetic refrigeration below $\sim$10~K is gadolinium gallium garnet (\ce{Gd3Ga5O12}, GGG), due to its high thermal conductivity and low lattice heat capacity and magnetic ordering temperature. \cite{pecharsky_magnetocaloric_1999} GGG only exhibits long-range ordering below $\sim$ 25 mK or in an applied field. Without an applied field, short-ranged ordering is observed, likely due to geometric frustration -- the magnetic Gd ions lie on two interpenetrating sub-lattices of corner-sharing triangles. \cite{petrenko_magnetic_1999} The lanthanide formates \ce{\textit{Ln}(HCO2)3} are another candidate magnetocaloric material. \ce{Gd(HCO2)3} is promising at $\sim$ 2 K while \ce{Tb(HCO2)3} has a larger magnetic entropy change above 4 K. A Gd-rich \ce{Gd_{1-x}Tb_x(HCO2)3} structure reportedly mixes the advantages of both compounds. \cite{saines_searching_2015}

In this investigation, the bulk magnetic properties of the \ce{Mg\textit{Ln}B5O10} compounds were measured for different lanthanides (\textit{Ln} = Pr, Nd, Sm--Er). A solution-based method was employed to achieve significantly higher sample purity than any ceramic (solid-state) synthesis approach. The \ce{Mg\textit{Ln}B5O10} compounds were found to be candidate materials for quasi-1D magnetism. Determination of the magnetocaloric entropy change showed that \ce{MgGdB5O10} in particular shows promise for application to solid state magnetic refrigeration in the liquid helium temperature range ($\geq$ 2 K).

\section{Methods}\label{Methods}

\subsection{Solution-based synthesis}

Our synthesis follows Zhang \textit{et al.} \cite{zhang_phase_2017} who reported phase-pure synthesis of \ce{MgYB5O10} \textit{via} the Pechini method, “which effectively suppressed the formation of \ce{YBO3} impurities." The solid reagents (\ce{\textit{Ln}2O3}, plus \ce{H3BO3} and \ce{MgO} each in 2 \%\ molar excess) were dissolved in dilute nitric acid and 1 molar equivalent of polyvinyl alcohol (PVA) was added to chelate the metal ions. (Reagent suppliers and purities are provided in the Supplemental Material.) The solution was stirred and heated at $\sim$ 90 $^{\circ}$C  until the liquid evaporated to form a gel, which was then ground, transferred to an alumina crucible and heated in a furnace to 750 $^{\circ}$C in air for 48 hours.

\subsection{Structural characterisation}
Powder X-Ray Diffraction (PXRD) was carried out at the I11 Beamline at the Diamond Light Source \cite{Diamond} with 15 keV X-rays, wavelength $\approx0.83$ \r{A}.\footnote{The wavelength was obtained by measuring a silicon standard. The refined values were 0.824813 \r{A} for the data collected in April 2025 (\textit{Ln} = Gd--Ho) and 0.826455 \r{A} for the data collected in July 2025 (\textit{Ln} = La, Pr, Nd, Sm, Eu, Er).} Powder samples were packed in 0.5 mm diameter borosilicate glass capillaries and sealed with epoxy glue. Rietveld refinements \cite{rietveld_profile_1969} were carried out using TOPAS software. \cite{coelho_topas_2018}

The PXRD pattern background was fitted to a Chebyshev polynomial with 18 terms. PXRD peak shape contributions from the beam were fitted to a Thompson-Cox-Hastings pseudo-Voigt peak shape. \cite{thompson_rietveld_1987} Sample contributions to peak shapes were accounted for where necessary using Lorentzian and Gaussian crystallite size parameters, and Stephens monoclinic strain broadening. \cite{Stephens}

\subsection{Magnetic characterisation}
Magnetic susceptibility and isothermal magnetisation were measured using a Quantum Design Magnetic Properties Measurement System (MPMS) with a Superconducting Quantum Interference Device (SQUID) magnetometer, in the temperature and field ranges $1.8\le T(K)\le300$ and $0\le \mu_0H(T) \le 7$ respectively. The susceptibility was measured on warming at a field of 500 Oe in both the zero-field-cooled (ZFC) and field-cooled (FC) regimes. The isothermal magnetisation was measured between 0 and 7 T at temperatures of 2, 4, 6, 8, 10, and 100 K. 

Magnetic susceptibility was calculated assuming that magnetisation varies linearly with field: $\chi = \partial M/\partial H \approx M/H$. Isothermal magnetisation data (figure~\ref{fig:Isotherm_mag}) show that the magnetisation is linear with field at low field strengths for all \textit{Ln} and all temperatures measured.

To determine the strength of the magnetocaloric effect, the magnetic entropy change was calculated from the measured sample isothermal magnetisation via: \cite{balli_advanced_2017,mukherjee_magnetic_2018,romero_gomez_magnetocaloric_2013,pecharsky_magnetocaloric_1999}

\begin{equation}
    \Delta S_m = \int^{H_f}_{H_i} \left. \frac{\partial M(H')}{\partial T} \right|_{p,H'} dH'
    \label{eqn:MCE}
\end{equation}

\subsection{Heat capacity}
Heat capacity measurements on \ce{MgTbB5O10} at 0, 0.5, 1, 2, 4 and 9 T were performed using a Quantum Design Physical Properties Measurement System (PPMS) in the temperature range 1.8 $\le$ T(K) $\le$ 30. To improve thermal conductivity at low temperatures, \ce{MgTbB5O10} was mixed with an approximately equal mass of silver powder (Alfa Aesar, 99.99\%, --635 mesh). The silver contribution to the total heat capacity was then deducted, using literature data \cite{smith_low-temperature_1995}. The lattice contribution to the \ce{Mg\textit{Ln}B5O10} heat capacity ($C_\mathrm{latt}$) was then subtracted via a Debye model with Debye temperature $\theta_\mathrm{D} = 224(1)$ K, leaving the magnetic contribution $C_\mathrm{mag}$.

\section{Results and Discussion}\label{Results}

\subsection{Sample Synthesis}\label{Synthesis}
Initially, a number of synthesis routes were attempted for the preparation of \ce{Mg\textit{Ln}B5O10} using a variety of reagents. \textit{Ln}-containing magnetic impurities (principally different forms of lanthanide borates) were present at $\ge$10 wt \%\ in every solid-state attempt. (Further details are given in the Supplemental Material.) Knitel \textit{et al.} \cite{knitel_photoluminescence_2000} report the same impurities in their solid-state synthesis attempts of \ce{Mg(La/Y)B5O10}, finding that ``neither variation of the excess \ce{H3BO3} nor prolonging of the firing time enhanced the purity of [their] samples." In this work, samples with $>95\%$ phase purity were achieved using a solution-based method, starting from lanthanide and magnesium oxides and PVA \cite{zhang_phase_2017}.

Phase-pure \ce{MgEuB5O10} was produced via the sol-gel method. For \textit{Ln} = Gd--Er, the dominant impurity was \ce{LnBO3} ($C2/c$). For the samples with larger lanthanides, La, Pr and Sm, the impurity phase was \ce{\textit{Ln}BO3} ($P2_1/m$). The impurity in the Nd sample could not be determined, but a comparison of the relative primary and secondary phase peak intensities between different compounds indicates that the Nd impurity is likely present at less than 2 wt \%. Finally, we note that neither solid-state nor solution-based synthesis succeeded in producing any \ce{MgYbB5O10}, instead producing various \ce{YbBO3} phases. This is in accordance with Saubat \textit{et al.} \cite{saubat_synthesis_1980} and it is therefore likely that the \ce{Mg\textit{Ln}B5O10} structure is unstable for \textit{Ln} ionic radii smaller than that of Er. 

\subsection{Structural Characterisation} \label{Lattice Parameters}
Samples were characterised using powder X-ray diffraction (PXRD) refined against the reported crystal structure \cite{zhang_phase_2017} using the Rietveld method. The background coefficients, lattice parameters, zero-point error, and the atomic coordinates of the lanthanide ion were refined, while the positions of the lighter Mg, B and O ions were fixed at values from literature reports. Fractional site occupancy values were fixed at 1 and isotropic thermal parameters fixed at 1~\r{A}$^2$ for all atoms.

The \ce{Mg\textit{Ln}B5O10} unit cell can be described by either the $P2_1/c$ or $P2_1/n$ settings of the same space group (number 14). $P2_1/c$ is the standard setting but $P2_1/n$ has a less oblique unit cell, with angle $\beta$ closer to 90$^\circ$. \cite{hahn_230_2002} As explained in detail elsewhere, \cite{Space_groups} using the less oblique cell reduces correlations between refined parameters, potentially improving fit accuracy. However, refinements with both cell settings were conducted. A lower $\chi^2$ and $R_{wp}$ were obtained with the $P2_1/c$ cell (keeping other parameters such as the background the same), so the structures are therefore reported for the $P2_1/c$ cell.

There was some difficulty achieving accurate fitted intensities of the $(100)$ and/or $(10\bar{2})$ reflections for some samples. The Stephens' monoclinic model for anisotropic strain \cite{Stephens} was found to improve the fits for those samples. 

\begin{figure}[htbp]
    \centering
    \includegraphics[width=1.0\linewidth]{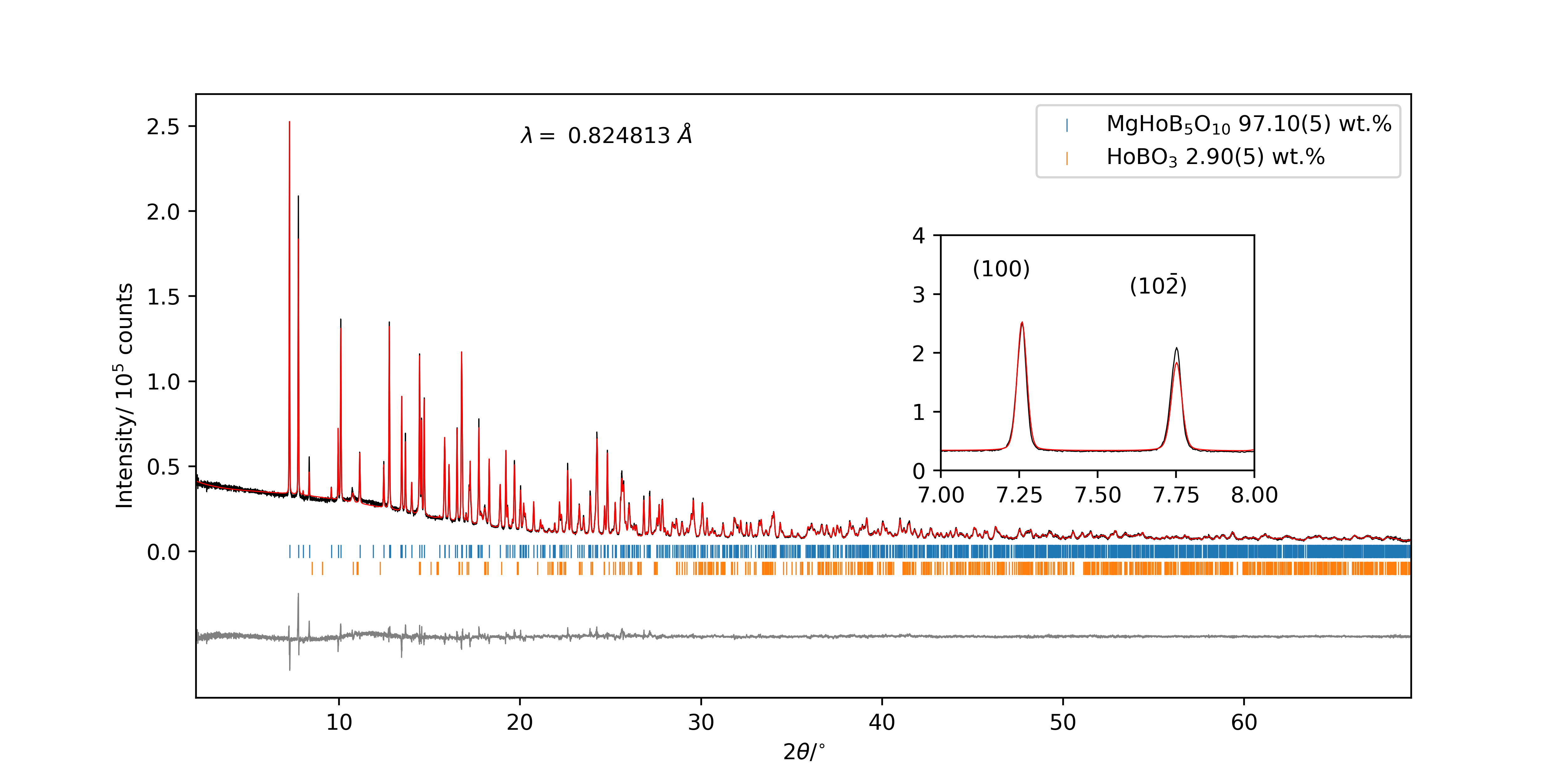}
    \caption{Rietveld refinement of synchrotron PXRD data \cite{Diamond} for \ce{MgHoB5O10}. Measured intensity is in black, calculated intensity in red and the difference offset below in grey. Inset: the $(100)$ and $(10\bar{2})$ reflections, illustrating the difficulty in achieving accurate fitted intensities. Refinement performed using TOPAS Software. \cite{coelho_topas_2018}}
    \label{fig:XRD}
\end{figure}

\begin{figure}[htbp]
    \centering
    \includegraphics[width=0.8\linewidth]{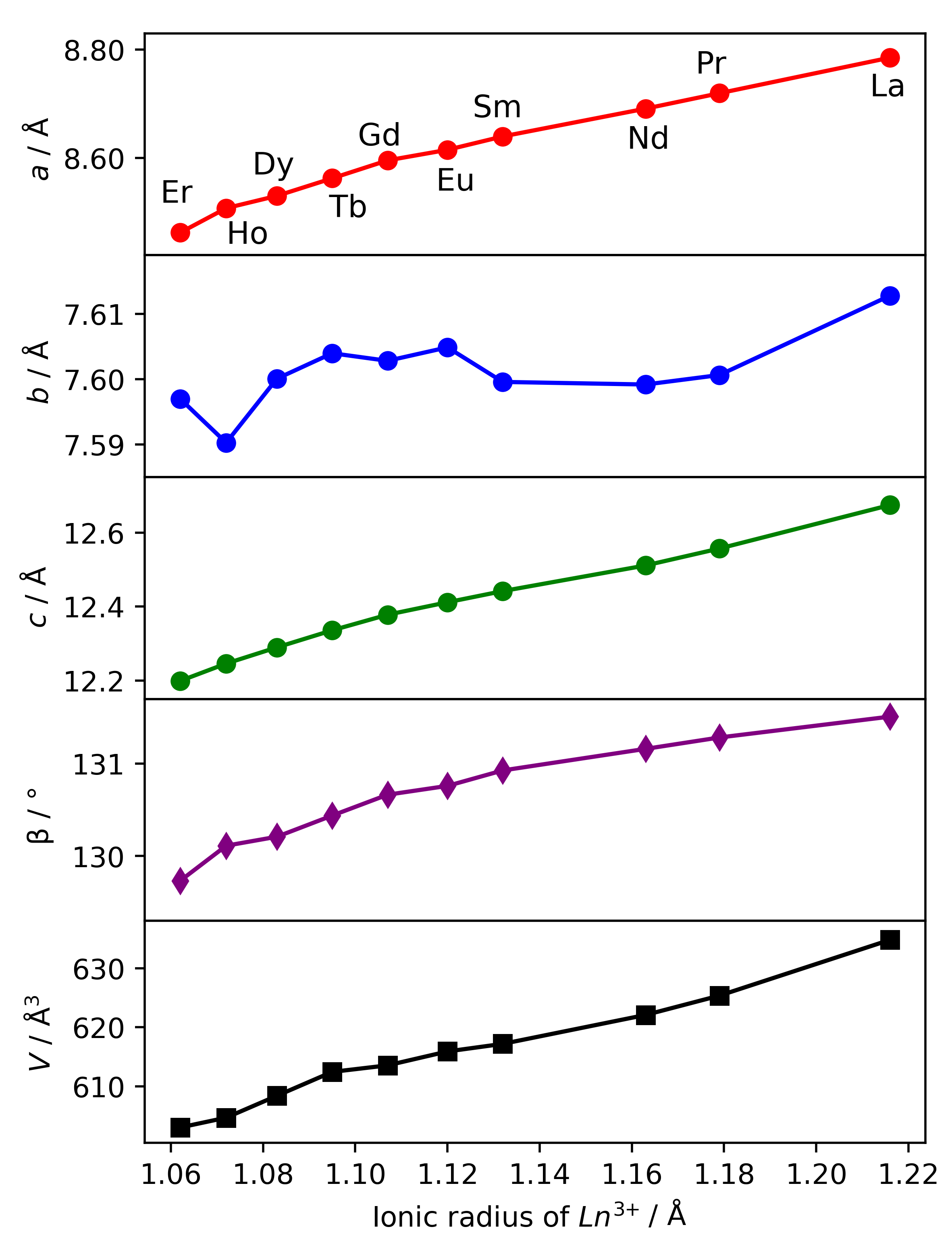}
    \caption{\ce{Mg\textit{Ln}B5O10} lattice parameters (obtained from Rietveld refinement of synchrotron PXRD data) \cite{Diamond} plotted against effective ionic radii for \textit{Ln} = (La,Pr,Nd,Sm--Er). Radii are for 9-coordinate $Ln^{3+}$ ions \cite{Shannon1976}. (Error bars are smaller than data points.)}
    \label{fig:lat}
\end{figure}

Figure~\ref{fig:XRD} shows a representative Rietveld refinement of PXRD data for \ce{Mg\textit{Ln}B5O10}. Refined lattice parameters for samples across the \textit{Ln} series are given in table~\ref{tab:lat}. The unit cell dimensions increase roughly linearly with effective lanthanide ionic radius, figure~\ref{fig:lat}.

\begin{landscape}
\begin{table}[htbp]
    \centering
    \begin{adjustbox}{width=1.5\textwidth}
    {\renewcommand{\arraystretch}{2}%
    \begin{tabular}{ccccccccccc}
     \hline
     \ce{Mg\textit{Ln}B5O10}&La&Pr&Nd&Sm&Eu&Gd&Tb&Dy&Ho&Er \\
    \hline
    Impurities, wt.$\%$&\ce{LaBO3}, 2.89(6)&\ce{PrBO3}, 2.69(6)&Unknown, $\sim$2&\ce{SmBO3}, 3.0(2)&None&\ce{GdBO3}, 2.14(5)&\ce{TbBO3}, 2.08(6)&\ce{DyBO3}, 2.47(5)&\ce{HoBO3}, 2.90(5)&\ce{ErBO3}, 6.35(10) \\
    \hline
    $a$/\r{A}&8.78519(6)&8.71942(5)&8.69097(6)&8.63959(6)&8.61484(8)&8.59528(11)&8.56220(7)&8.52971(6)&8.50700(7)&8.46170(8) \\
    \hline
    $b$/\r{A}&7.61281(5)&7.60064(4)&7.59919(4)&7.59959(5)&7.60488(6)&7.60282(10)&7.60398(6)&7.60008(6)&7.59027(7)&7.59701(6) \\
    \hline
    $c$/\r{A}&12.67441(9)&12.55682(7)&12.51107(8)&12.44167(8)&12.41147(11)&12.37778(16)&12.33565(10)&12.28948(9)&12.24570(11)&12.19863(10) \\
    \hline
    $\beta$/$^{\circ}$&131.5088(4)&131.2830(3)&131.160(3)&130.9267(3)&130.7594(5)&130.6639(7)&130.4385(6)&130.2092(6)&130.1123(6)&129.7322(5) \\
    \hline
    $V$/\r{A}$^3$&634.777(8)&625.350(7)&622.089(7)&617.197(8)&615.916(10)&613.563(15)&611.268(10)&608.421(10)&604.722(10)&603.061(10) \\
    \hline
    \textit{Ln}$(x,y,z)$&\makecell{0.05257(13)\\0.81280(9)\\0.23518(8)}&\makecell{0.05188(10)\\0.81353(7)\\0.23615(7)}&\makecell{0.05169(12)\\0.81302(8)\\0.23601(8)}&\makecell{0.05091(10)\\0.81389(7)\\0.23664(7)}&\makecell{0.05147(14)\\0.18455(9)\\0.23754(9)}&\makecell{0.05049(14)\\0.81421(9)\\0.23732(9)}&\makecell{0.05192(13)\\0.81475(8)\\0.23813(8)}&\makecell{0.05265(12)\\0.81598(9)\\0.23817(8)}&\makecell{0.05378(11)\\0.81516(9)\\0.23894(8)}&\makecell{0.05670(13)\\0.81621(9)\\0.24164(9)} \\
    \hline
    \makecell{\textit{Ln-Ln} $R_\mathrm{intra}$ /\r{A}}&3.9934(12)&3.9746(7)&3.9725(8)&3.9631(7)&3.9622(9)&3.9570(7)&3.9584(7)&3.9584(8)&3.9539(6)&3.9549(8) \\
    \hline
    \makecell{\textit{Ln-Ln} $R_\mathrm{inter}$ /\r{A}}&6.101(4)&6.0782(19)&6.064(3)&6.0549(19)&6.057(3)&6.0501(19)&6.0388(19)&6.009(3)&6.0047(18)&6.020(3) \\
    \hline
    $\chi^2$&1.466&1.352&1.463&1.330&1.250&3.264&3.277&3.020&2.821&1.673 \\
    \hline
    $R_{wp}$&5.113&4.741&5.512&4.718&4.395&3.762&3.808&4.242&3.642&4.196 \\
    \hline
    \end{tabular}%
    } \quad
    \end{adjustbox}
    \caption{Lattice parameters and sample impurity compositions for \ce{Mg\textit{Ln}B5O10} as obtained from room temperature PXRD. \textit{Ln--Ln} nearest neighbour and next-nearest neighbour bond lengths were determined in VESTA. \cite{momma_vesta_2011}}
    \label{tab:lat}
\end{table}
\end{landscape}

\subsection{Magnetic Susceptibility}
\label{Susceptibility}
\begin{figure}[htbp]
    \vspace*{-2.0in}
    \centering
    \includegraphics[width=0.85\textwidth]{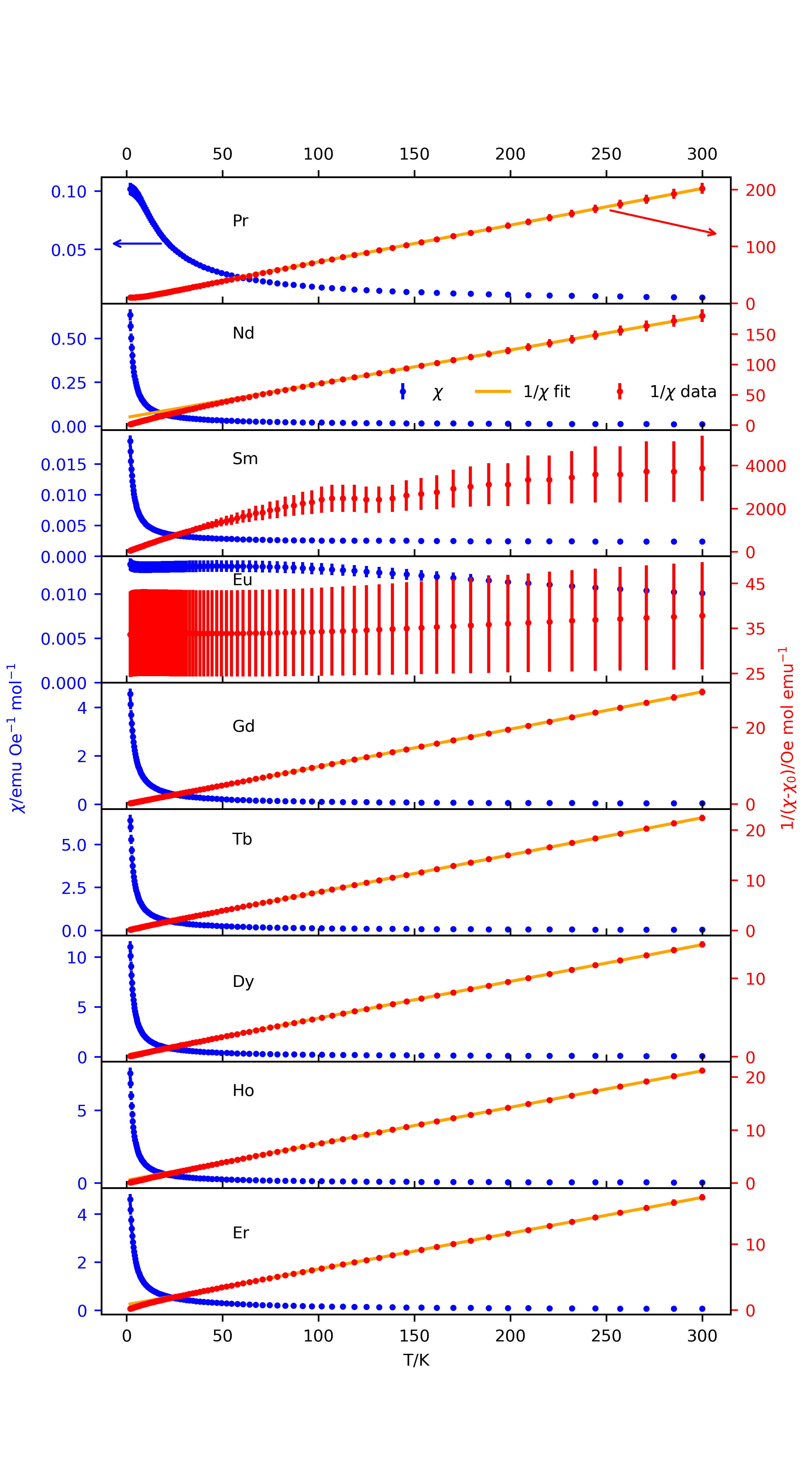}
    \caption{Magnetic susceptibility $\chi$ vs T (blue), and $1/(\chi-\chi_0)$ vs T. ($1.8\leq T(\mathrm{K}) \leq 300$; data in red, Curie-Weiss fit in orange.) Linear fits were obtained down to $\sim$80 K (and lower for Gd). Sm and Eu did not obey the Curie-Weiss Law.}
    \label{fig:susceptibilities}
\end{figure}

The zero-field-cooled (ZFC) magnetic susceptibilities $\chi(T)$ in 500 Oe for \ce{Mg\textit{Ln}B5O10} samples between 1.8 and 300 K are plotted in figure~\ref{fig:susceptibilities}. The magnetic moments were particularly small for \ce{MgEuB5O10}, hence the relatively large error bars, which reflect the error in the sample mass ($\pm$ 0.2 mg). 

\begin{figure}[htbp]
    \centering
    \includegraphics[width=0.9\linewidth]{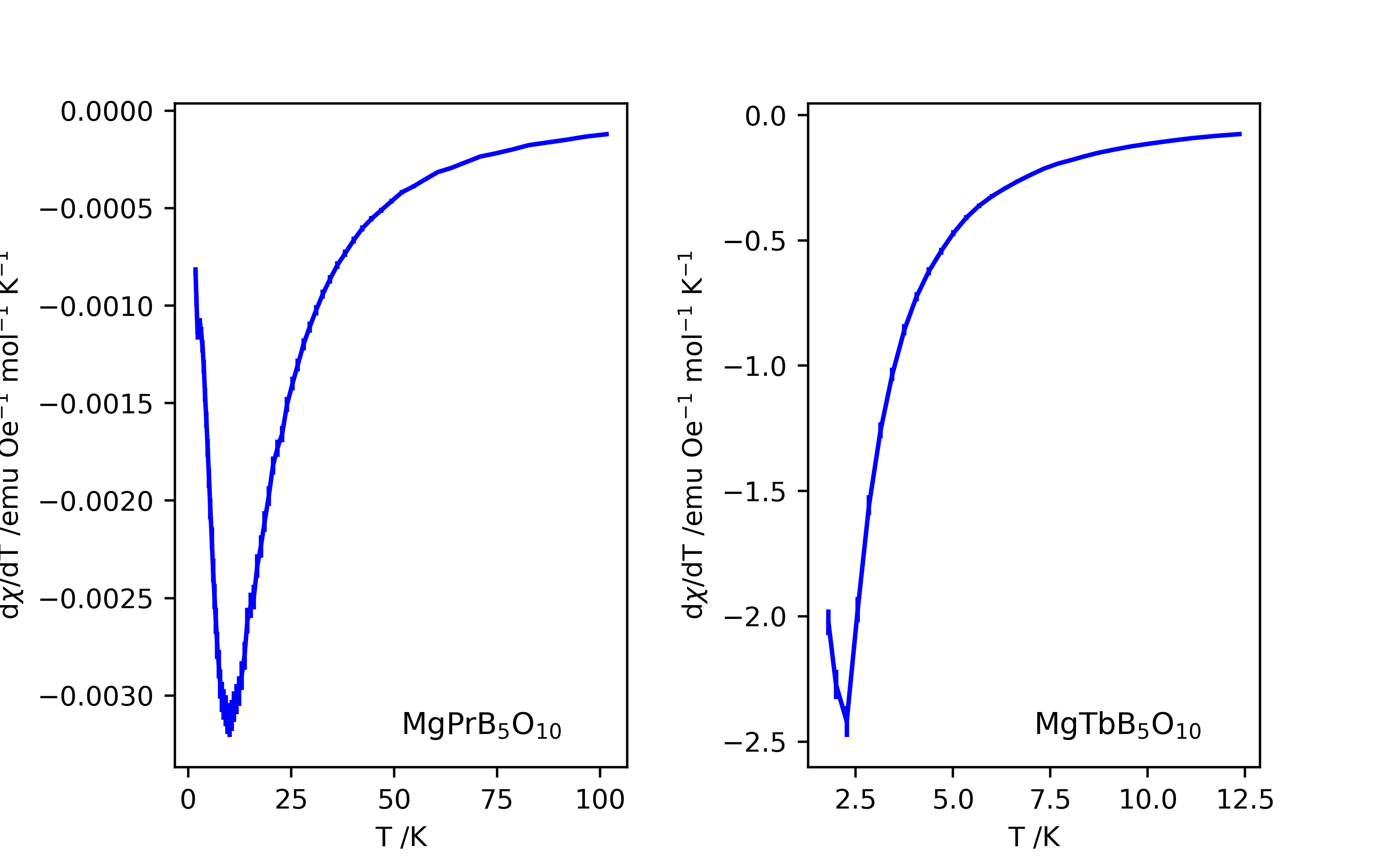}
    \caption{$d\chi/dT$ vs temperature for \ce{MgPrB5O10} (left) and \ce{MgTbB5O10} (right). Sharp changes in $d\chi/dT$ at $\sim$10 and 2.25 K respectively indicate the onset of an ordering transition.}
    \label{fig:dchi_dT}
\end{figure}

For most \textit{Ln}, no ordering was observed down to 1.8 K. However, \ce{MgPrB5O10} and \ce{MgTbB5O10} showed signs of ordering at $\sim$10 and 2.25 K respectively: $d\chi/dT$ reaches a minimum at these temperatures (figure~\ref{fig:dchi_dT}). The broadened $\chi(T)$ shape for \ce{MgPrB5O10} may be indicative of a singlet ground state, which is typical for Pr (a non-Kramers ion) in low-symmetry crystal structures: Similar features have been reported in Pr compounds, including the quasi-1D \ce{Ca4PrO(BO3)3} \cite{kelly_magnetic_2020} and \ce{Pr(BO2)3} \cite{mukherjee_magnetic_2017}. We note that the site symmetry of Pr in \ce{MgPrB5O10} \cite{saubat_synthesis_1980} is not sufficiently high symmetry for a protected doublet ground state, supporting this hypothesis.

The ordering temperature observed for the Tb sample ($\sim$2.25 K) is close to that reported for short-range ordering in \ce{TbBO3} (2.02 K) \cite{mukherjee_magnetic_2018}. However, the impurity is present at only 2.08(6) wt $\%$ in the sample. Therefore, the clear cusp in Fig.~\ref{fig:susceptibilities} is very unlikely to come from \ce{TbBO3}, especially given our subsequent heat capacity data (see Section~\ref{section:HC}).

The temperature-dependent magnetic susceptibility of \ce{MgEuB5O10} resembles that reported for \ce{EuBO3}. \cite{laureiro_synthesis_1991} The spacings of the $^7F_J$ multiplet energy levels are small compared to $k_B T$, so not all \ce{Eu^{3+}} are in the ground state, resulting in a strong van Vleck contribution to the susceptibility. Excited states also contribute to $\chi$ in Sm.

Also shown in figure~\ref{fig:susceptibilities} are the inverse susceptibilities $1/(\chi - \chi_0)$ and high temperature fits to the Curie-Weiss Law (\ref{eqn:CVL2}):

\begin{equation}
    \chi(T) = \chi_0 + \frac{C}{T-\theta_\mathrm{CW}}
    \label{eqn:CVL2}
\end{equation}

where $C$ is the Curie constant and $\theta_\mathrm{CW}$ the Curie-Weiss temperature. Good linear fits were obtained for most \textit{Ln} in the temperature range T$\ge$ 80 K. For Gd, fitting was possible for T $\gtrsim$ 25 K. \ce{MgSmB5O10} appears to have a Curie-Weiss-like $\chi(T)$, but the linear fit was poor even at higher temperatures, possibly due to a van Vleck contribution to the susceptibility. \cite{Sanders2017} 

Parameters from the Curie-Weiss fits are presented in table~\ref{tab:CW}. For all samples $\theta_\mathrm{CW}$ is negative, indicative of antiferromagnetic correlations. The effective magnetic moments $\mu_\mathrm{eff} = \sqrt{8C}\mu_\mathrm{B}$ are reasonably consistent with the free ion values $\mu_\mathrm{eff} = g_J \sqrt{J(J+1)}$. Field-cooled susceptibility data matched the ZFC data very closely. 

\begin{table}[htbp]
    \centering
    \resizebox{\textwidth}{!}{%
    \begin{tabular}{ccccccccc}
    \hline
    \textit{Ln}&$\chi_0$/emu~Oe$^{-1}$~mol$^{-1}$&$\theta_\mathrm{CW}$/K&$\mu_\mathrm{eff}/\mu_\mathrm{B}$&Theoretical $\mu_\mathrm{eff}/\mu_\mathrm{B}$&$f=|\theta_\mathrm{CW}|/T_c$ (*) &$J_1$/K&$D_\mathrm{intra}$/K&$D_\mathrm{inter}$/K \\
    \hline
    Pr&0.00373(5)&-13.2(2)&3.37(2)&3.58&1.31(4)&0.493(7)&0.0564(7)&0.00158(2) \\
    Nd&0.00539(6)&-22.9(9)&3.38(2)&3.62&$>$12.7(5)&0.695(5)&0.00459(5)&0.001291(14) \\
    Gd&0.00512(5)&-0.50(1)&7.91(4)&7.94&$>$0.277(6)&0.024(8)&0.0400(4)&0.01118(15) \\
    Tb&0.00251(3)&-3.4(1)&9.71(6)&9.72&1.51(4)&0.061(3)&0.0225(3)&0.00635(7) \\
    Dy&0.00549(5)&-4.97(12)&10.41(4)&10.65&$>$2.76(7)&0.058(2)&0.01708(14)&0.00488(4) \\
    Ho&0.00239(3)&-7.7(2)&10.78(7)&10.61&$>$4.3(1)&0.080(2)&0.0163(2)&0.00464(6) \\
    Er&0.00527(5)&-15.2(3)&9.94(4)&9.58&$>$8.46(16)&0.179(2)&0.01561(13)&0.00443(4) \\
    \hline
    \end{tabular}%
    }
    \caption{Curie-Weiss fitted parameters and derived quantities for \ce{Mg\textit{Ln}B5O10} samples. (*) Where no ordering is observed, we used $T_c<1.8$ to estimate a lower bound for $f$.}
    \label{tab:CW}
\end{table}

In regular non-frustrated systems, $\theta_\mathrm{CW}$ and the ordering temperature $T_c$ are similar in magnitude. In frustrated systems, ordering, and hence $T_c$, are suppressed.
Ramirez introduces a frustration index $f = |\theta_\mathrm{CW}|/T_c$ to quantify the extent of this effect. \cite{ramirez_strongly_1994} $f\gtrsim10$ implies strong frustration. \cite{mukherjee_magnetic_2017} Values of the frustration index are presented in table~\ref{tab:CW} using the fitted $\theta_\mathrm{CW}$ values. Where ordering was not observed down to 1.8 K, $T_c < 1.8$~K provides a lower bound on $f$. \ce{MgNdB5O10} and \ce{MgErB5O10} appear to have the greatest degree of frustration in the absence of sub-Kelvin data, which may reveal lower ordering temperatures and hence greater frustration in other \textit{Ln}. Measurements at lower temperatures would be required to explore this further.

To be considered as quasi-1D, the intra-chain exchange ($J_\mathrm{intra}$, equation~\ref{eqn:J1}) and dipolar ($D$, equation~\ref{eqn:D}) interactions must exceed the inter-chain interactions. For through-bond exchange, the relevant pathways are \textit{Ln}-\ce{O}-\textit{Ln} and \textit{Ln}-\ce{O}-B-\ce{O}-\textit{Ln} respectively. We therefore anticipate $J_\mathrm{intra} >> J_\mathrm{inter}$. Dipolar interactions scale as $D \propto 1/R^3$ and so can be calculated in the mean field limit: 

\begin{equation}
    J_\mathrm{intra} \sim \frac{3 k_\mathrm{B} \theta_\mathrm{CW}}{2nJ(J+1)}
    \label{eqn:J1}
\end{equation}

\begin{equation}
    D \sim \frac{\mu_0 \mu_{eff}^2}{4 \pi R^3 J(J+1)}
    \label{eqn:D}
\end{equation}

($n$ is the number of nearest neighbours (2), $R$ the through-space $Ln^{3+}$ separation within/between the chains and $J$ the total angular momentum quantum number.) 

$J_\mathrm{intra}$ and $D_\mathrm{intra,inter}$ are listed in table~\ref{tab:CW}. These values assume mean field behaviour, which is only valid in Heisenberg systems and so should be treated with caution. Nevertheless, one can tentatively conclude that \ce{Mg(Pr,Nd,Tb,Dy,Ho,Er)B5O10} are candidate materials for quasi-1D magnetism since $J_\mathrm{intra} > D_\mathrm{intra} > D_\mathrm{inter} >> J_\mathrm{inter}$.

\subsection{Isothermal Magnetisation}
\label{Isomag}
\begin{figure}[htbp]
    \centering
    \includegraphics[width=0.9\textwidth]{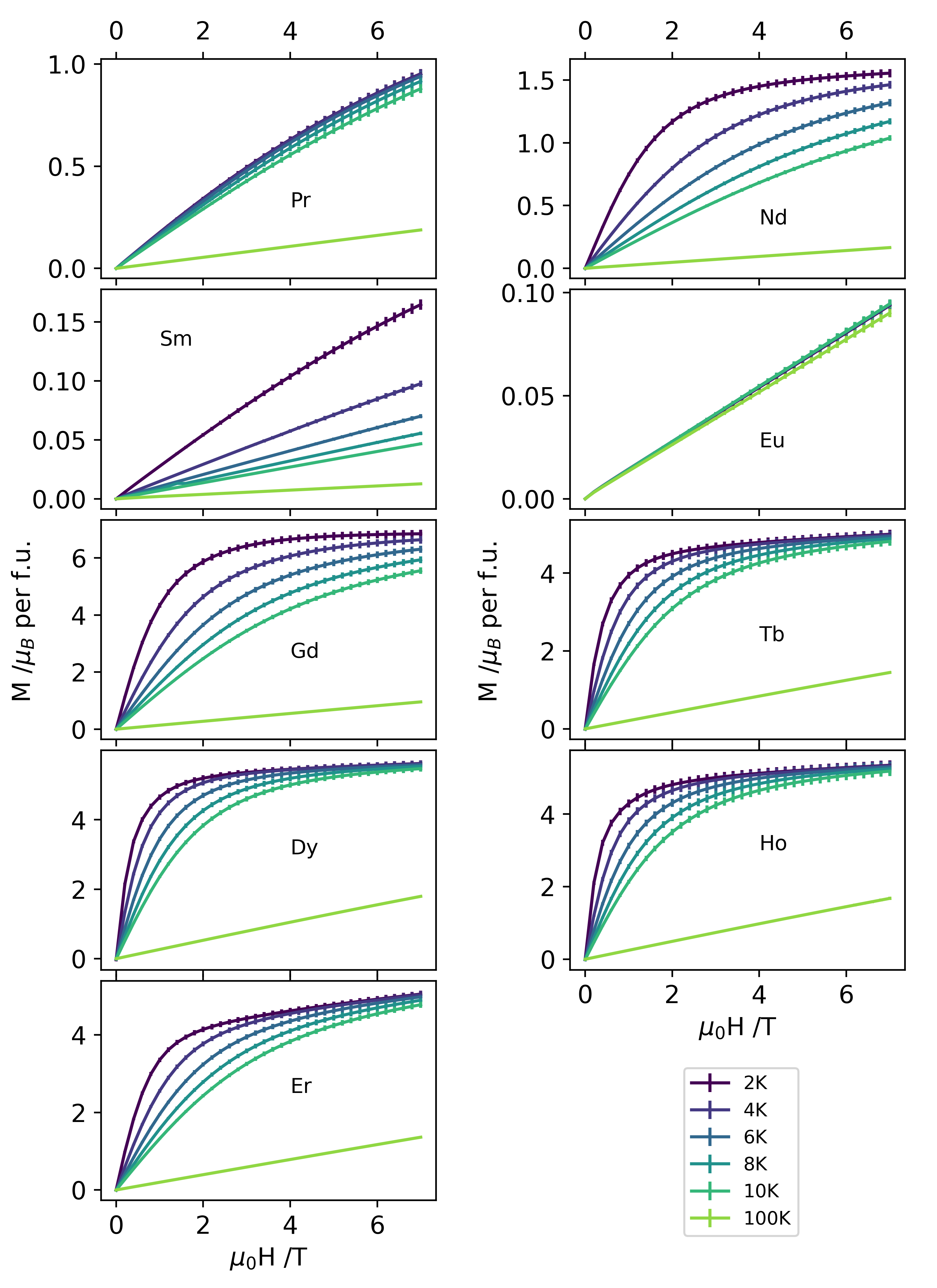}
    \caption{Isothermal magnetisation $M(H)/\mu_\mathrm{B}$ per formula unit for \ce{Mg\textit{Ln}B5O10} at 2, 4, 6, 8, 10 and 100 K between 0 and 7 T.}
    \label{fig:Isotherm_mag}
\end{figure}

Isothermal magnetisation $M(H)$ curves for \ce{Mg\textit{Ln}B5O10} are plotted in figure~\ref{fig:Isotherm_mag}.  

All samples are paramagnetic at 100 K ($M\propto H$). \textit{Ln} = Pr, Sm and Eu remain largely paramagnetic up to 7 T for each measured temperature. The remaining \textit{Ln} all show the onset of spin saturation.

To determine the nature of this saturation, $M(H)$ at 2 K for \textit{Ln} = (Nd, Gd-Er) were fitted (using the scipy.curvefit package) to the modified Brillouin function $B_J(y)$ (equation~\ref{eqn:Brill_mod}), assuming free spins at low temperature ($y = g_J \mu_\mathrm{B} J B/k_\mathrm{B} T$; $M_s$ is the saturation magnetisation). 

\begin{equation}
    M = M_s B_J(y) + \chi_1 H
    \label{eqn:Brill_mod}
\end{equation}

Here, the linear term $\chi_1 H$ incorporates any van Vleck paramagnetism. \cite{blundell_magnetism_2001} A linear fit was first applied between $\sim$4-7 T, to obtain an initial estimate of $\chi_1$ for the main fitting. Obtaining good fits to the modified Brillouin function proved challenging. (Further detail is given in the Supplemental Material.) Nonetheless, it provided estimates of the saturation magnetisation $M_s$, which permitted clear distinction between Heisenberg and Ising spin systems in the \ce{Mg\textit{Ln}B5O10} family.

Calculated estimates for the saturation magnetisation are presented in table~\ref{tab:Msat}, along with theoretical $M_s$ values.  
$M_s = \mu_\mathrm{B} g_J J$ is expected for Heisenberg spins while $M_s = \mu_\mathrm{B} g_J J/2$ indicates Ising spin saturation, implying single ion anisotropy. \cite{blundell_magnetism_2001} (Both $M_s$ equations are per formula unit.) \textit{Ln} = Nd, Tb-Er showed Ising-like spin saturation, while \ce{MgGdB5O10} showed Heisenberg spins.

\begin{table}[htbp]
    \centering
    \begin{tabular}{cccc}
     \hline
     \textit{Ln}&$M_s/\mu_\mathrm{B}$ per f.u.&$g_JJ$&$\chi_1/\mu_\mathrm{B}$ per f.u. T$^{-1}$ \\
     
     \hline
     Nd & 1.64(7) & 3.27 & 0 \\
     Gd & 6.6(5) & 7 & 0.05(5) \\
     Tb & 4.8(3) & 9 & 0.03(3)\\
     Dy & 5.5(3) & 10 & 0.03(3)\\
     Ho & 4.8(3) & 10 & 0.04(4)\\
     Er & 4.41(8) & 9 & 0.0937(14)\\
     \hline
    \end{tabular}
    \caption{Estimated saturation magnetisation for \ce{Mg\textit{Ln}B5O10}. These broadly agree with expected values: $g_J J$ for Heisenberg spins (Gd) and $g_J J/2$ for Ising spins (other \textit{Ln}). Errors in $\chi_1$ are large due to the poor fit quality. (See Supplementary Material.)}
    \label{tab:Msat}
\end{table}

A magnetisation plateau at $M_s/3$, implying ordering within but not across chains, is strongly suggestive of quasi-1D magnetism. \cite{mukherjee_magnetic_2017} This was not observed for any \ce{Mg\textit{Ln}B5O10} at the fields and temperatures investigated. However, the lack of clear evidence for 3D long range ordering means that quasi-1D behaviour may still be possible. More data at different temperatures and fields are needed to clarify the situation. 

\subsection{Heat Capacity} \label{section:HC}

\begin{figure}[htbp]
    \centering
    \includegraphics[width=1.0\linewidth]{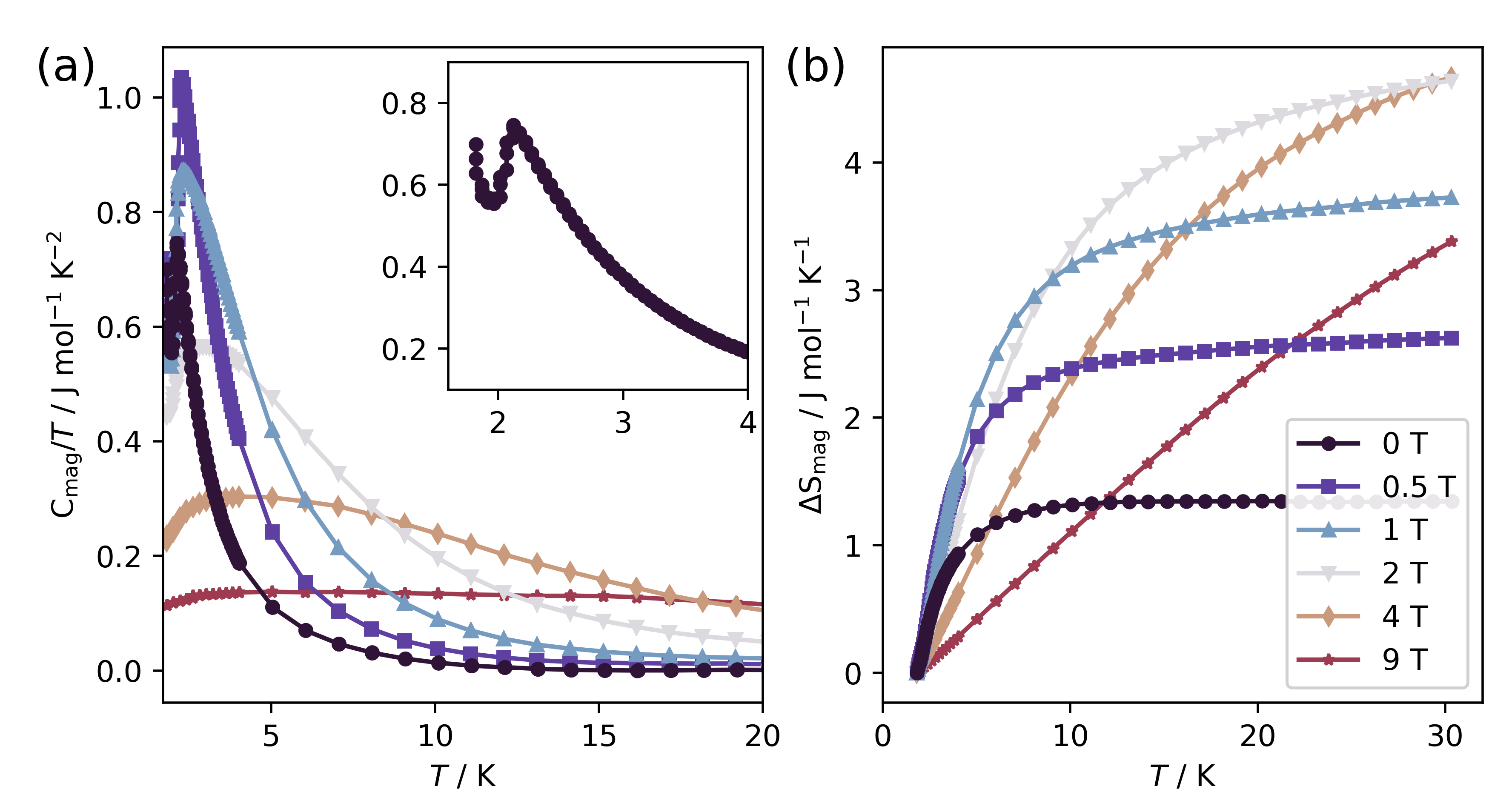}
    \caption{(a) Magnetic heat capacity $C_\mathrm{mag}/T$ for \ce{MgTbB5O10} in different applied fields. The inset shows a zoomed-in part of the zero-field data at low temperature. (b) Magnetic entropy change $\Delta S_\mathrm{mag}$ as a function of field and temperature.}
    \label{fig:HC}
\end{figure}

Since the susceptibility of \ce{MgTbB5O10} showed evidence of magnetic order above 2~K, we performed heat capacity measurements (1.8--30~K, 0--9~T) to investigate the transition further. The temperature-normalised magnetic heat capacity $C_\mathrm{mag}/T$ and the magnetic entropy change $\Delta S_\mathrm{mag}$ (obtained by integration of $C_\mathrm{mag}/T$) are plotted against temperature in figure~\ref{fig:HC}.

There is a clear peak in $C_\mathrm{mag}/T$ at 2.12~K at zero field (Fig.~\ref{fig:HC}(a), inset), which broadens and moves to higher temperatures as the field increases. The zero-field peak is somewhat broader than the sharp ``lambda-like'' transition expected for long-range 3D magnetic order as observed in other Tb compounds \cite{Kelly2022a,Hill1986}. Therefore, we postulate that the magnetic correlations might be short-ranged with onset temperature $T_\mathrm{o}\approx2.25$~K. Examining the magnetic entropy change, we see that it falls considerably short of the maximum expected value $R \ln{(2J+1)} =$ 21.32 J K$^{-1}$ mol$^{-1}$ for \ce{MgTbB5O10} with $J=6$. However, the maximum entropy change for an effective spin-$\frac{1}{2}$ (Ising) spin system (as deduced from our $M(H)$ data) would be $R \ln{(2)} =$ 5.76 J K$^{-1}$ mol$^{-1}$ and our observed maximum $\Delta S_\mathrm{mag}$ reaches 80~\%\ of $R \ln{2}$: the `missing' 20~\%\ is probably a systematic offset owing to the missing data below 1.8~K, and/or might relate to the suggested short magnetic correlation length. Future neutron diffraction and total scattering experiments would be necessary to determine the magnetic structure and correlation length respectively.

\subsection{Magnetocaloric Effect} \label{Magcol}

\begin{figure}[htbp]
    \centering
    \includegraphics[width=0.9\textwidth]{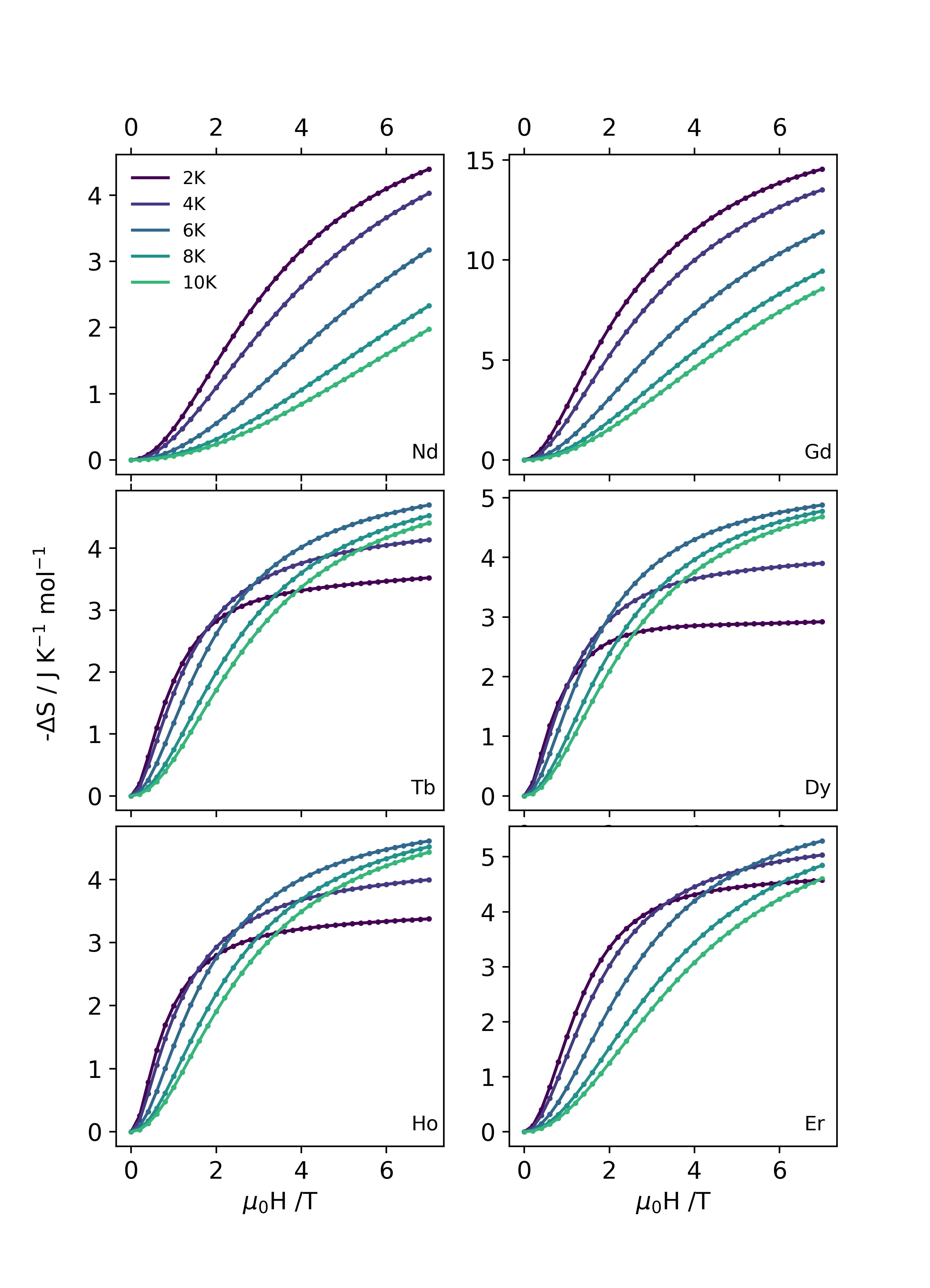}
    \caption{Magnetocaloric entropy change $|\Delta S|$ between 0 and 7 T at temperatures of (2,4,6,8,10) K.}
    \label{fig:magnetocalorics}
\end{figure}

\begin{figure}[htbp]
    \centering
    \includegraphics[width=0.5\linewidth]{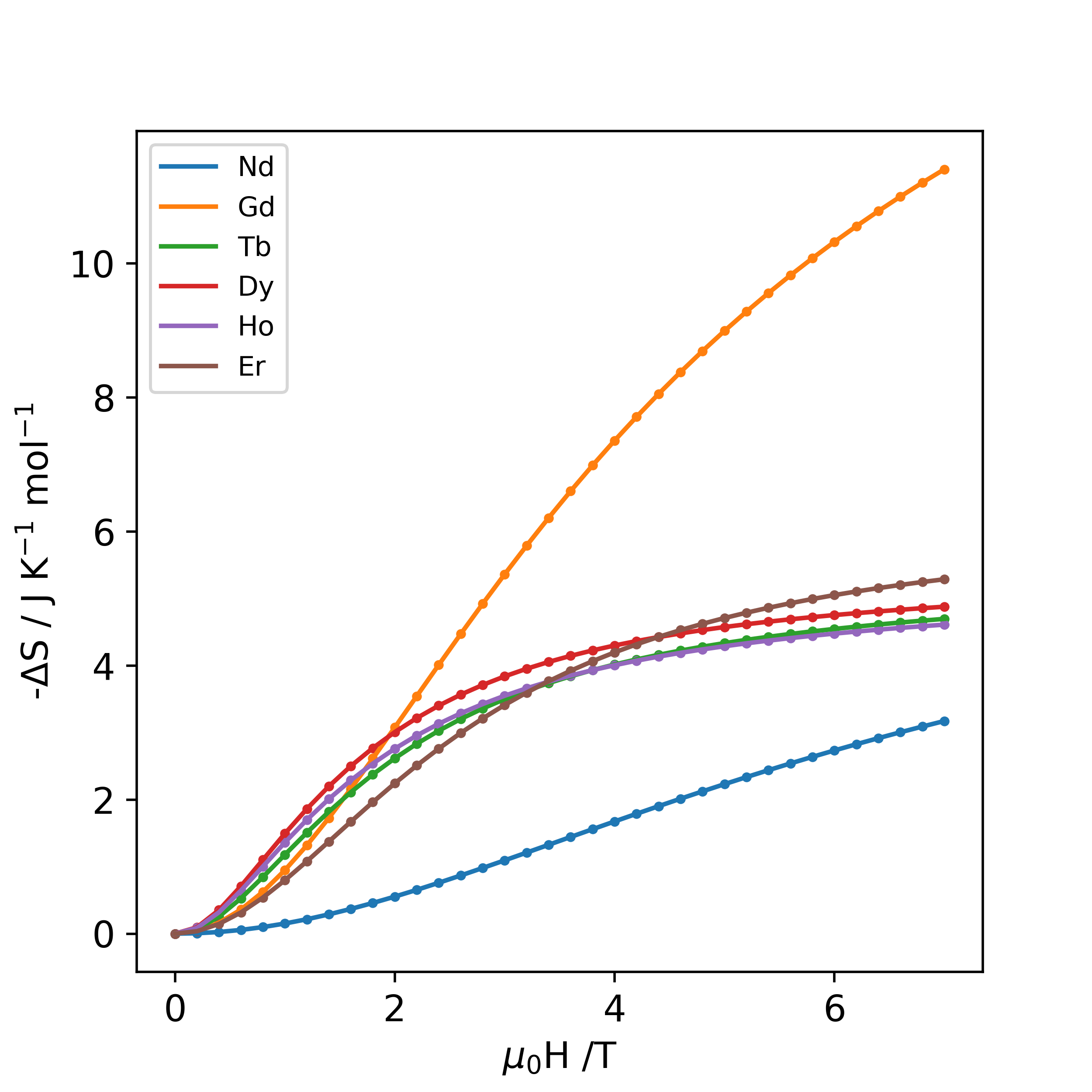}
    \caption{Negative magnetocaloric entropy change vs \textit{Ln} and field strength in \ce{Mg\textit{Ln}B5O10} at 6 K. \ce{MgDyB5O10} is superior in lower field strengths, while \ce{MgGdB5O10} far surpasses the other \textit{Ln} in large fields.}
    \label{fig:magcol_Ln}
\end{figure}

The magnetocaloric entropy change (equation \ref{eqn:MCE}) was calculated numerically for \textit{Ln} = Nd, (Gd--Er) using the isothermal magnetisation data and the resulting curves are plotted in figure~\ref{fig:magnetocalorics}. Typically, $|\Delta S|$ is greater at lower temperatures and higher field strengths. However, for \textit{Ln} = (Tb--Er), a field-induced transition is observed around 2--4 K, as evidenced by the overlapping curves in Fig.~\ref{fig:magnetocalorics}, meaning that the MCE in high fields is maximised at 6~K rather than at the lowest temperature. Future measurements of the susceptibility and/or heat capacity under different applied magnetic fields could be used to investigate these transitions further. Figure~\ref{fig:magcol_Ln} compares the MCE for different \textit{Ln} at 6~K, where the field-induced transition has not occurred. The MCE of \ce{MgGdB5O10} greatly outperforms all other \ce{Mg\textit{Ln}B5O10} at higher fields ($>2$~T), which is typical for Gd ($S=\frac{7}{2}$), whilst at lower fields it is overtaken by \ce{MgDyB5O10}. Similar behaviour was reported in the borates \ce{\textit{Ln}BO3} \cite{mukherjee_magnetic_2018}. This agrees with observations in previous work that Crystal Electric Field (CEF)-dominated materials with Heisenberg spin systems are more promising at high field while materials with substantial single-ion anisotropy perform better at low field. \cite{mukherjee_magnetic_2018,numazawa_magneto_2003}

The maximum $|\Delta S|$ observed, 14.5 J~K$^{-1}$~mol$^{-1}$ for \ce{MgGdB5O10} at 2~K and 7~T, corresponds to $\sim$84$\%$ of the maximum $R \ln{(2J+1)} \approx$ 17.3 J~K$^{-1}$~mol$^{-1}$ for free Heisenberg spins. Comparisons with other promising magnetocaloric materials in the literature at 2 K are presented in table~\ref{tab:magcol_comp}. The MCE for \ce{MgGdB5O10} at 7~T is comparable to gadolinium gallium garnet (GGG) at 9~T in both molar and mass units. \ce{MgGdB5O10} also performs strongly at 2~T, being comparable to \ce{Gd(HCO2)3} and surpassing dysprosium gallium garnet (DyGG) and \ce{DyBO3} -- which is unusual as \ce{MgGdB5O10} is a Heisenberg spins system. \ce{MgGdB5O10} is therefore a promising magnetocaloric material in both high and low field regimes. The other \ce{Mg\textit{Ln}B5O10} have similar entropy changes to their respective \ce{\textit{Ln}BO3}, but the field-induced transition renders them more suitable above $\sim$6 K. This could be useful for applications in the 4--10 K range, such as in cooling the niobium-based superconducting magnets for magnetic resonance imaging (MRI) scanners, to complement other paramagnetic magnetocaloric materials which are optimised for temperatures below 4 K. \cite{saines_searching_2015}

\begin{table}[htbp]
    \centering
    \begin{tabular}{cccc}
    \hline
    Compound&Field/T&$|\Delta S|$/J K$^{-1}$ mol$^{-1}_{Ln}$&$|\Delta S|$/J K$^{-1}$ kg$^{-1}$ \\
    \hline
    \ce{MgGdB5O10}&2&6.7&16.9 \\
     &7&14.5&36.7 \\
    \hline
    \ce{MgDyB5O10}&2&2.6&6.5 \\
    \ce{MgErB5O10}&2&3.3&8.1 \\
    \ce{MgTbB5O10}&2&2.8&7.0 \\
    \hline
    GGG \cite{mukherjee_sensitivity_2017} &9&14.1&41.8 \\
    DyGG \cite{mukherjee_sensitivity_2017} &2&3.8&11.1 \\
    \hline
    \ce{GdBO3} \cite{mukherjee_magnetic_2018} &9&12.5&57.8 \\
    \ce{DyBO3} \cite{mukherjee_magnetic_2018} &2&3.1&14.0 \\
    \ce{ErBO3} \cite{mukherjee_magnetic_2018} &2&2.6&11.5 \\
    \hline
    \ce{Gd(HCO2)3} \cite{saines_searching_2015} &2&6.1&20.9 \\
    \hline
    \end{tabular}
    \caption{Magnetocaloric entropy changes of selected materials at 2 K.}
    \label{tab:magcol_comp}
\end{table}

\section{Conclusions} \label{Conclusion}

High-purity samples of \ce{Mg\textit{Ln}B5O10} were obtained using a solution-based process and their crystal structures analysed using high-resolution powder diffraction. Magnetic susceptibility measurements were carried out on nine compounds in the series. \ce{MgSmB5O10} and \ce{MgEuB5O10} exhibited significant van Vleck paramagnetism, while \ce{MgPrB5O10} showed evidence for a singlet ground state. The other samples ($Ln=$ Nd, Gd, Tb, Dy, Ho, Er) have antiferromagnetic spin correlations, but no ordering above 2~K except for \ce{MgTbB5O10}, which showed signs of magnetic ordering at $\sim2.25$~K in both susceptibility and heat capacity data.

Isothermal magnetisation measurements revealed that \ce{Mg\textit{Ln}B5O10} ($Ln=$ Nd, Tb, Dy, Ho, Er) show Ising-like behaviour, while \ce{MgGdB5O10} has a Heisenberg spin system. The magnetocaloric entropy change was calculated for $Ln =$ (Nd, Tb--Er), with \ce{MgGdB5O10} in particular showing promise for application to solid-state cryogenic refrigeration. Several samples underwent a field-induced transition around 2--4 K at fields above $\sim2$ T. Future heat capacity measurements under an applied magnetic field could shed light on this transition. Finally, comparison of the strength of the exchange and dipole interactions suggests that many of the magnesium lanthanide borates remain candidates for quasi-1D magnetism. The \ce{Mg\textit{Ln}B5O10} system displays a range of interesting magnetic behaviour, and it is hoped that this investigation will motivate further study of these materials. 

\section*{Acknowledgements}
We acknowledge funding from the EPSRC for the use of the Advanced Materials Characterisation Suite (EP/M000524/1). N.D.K.~acknowledges funding from Jesus College, Cambridge for a Research Fellowship. We acknowledge I11 beamline at the Diamond Light Source, UK, for the synchrotron XRD measurements done under BAG proposal CY34243, and thank Farheen N.~Sayed for organising the beamtime. 

Data associated with this publication are available in the Cambridge University Repository (doi:10.17863/CAM.122845).

\section*{CRediT Author Contributions}
LGMR: Investigation, Formal analysis, Visualization, Writing - original draft. SED: Funding acquisition, Resources, Supervision, Writing - review and editing. NDK: Conceptualization, Investigation, Formal analysis, Supervision, Writing - review and editing. 

\bibliographystyle{elsarticle-num} 
\bibliography{editable_refs}

\newpage

\appendix{
    \begin{center}
    \textbf{SUPPLEMENTAL MATERIAL: Magnetic Properties of the Quasi-1D Magnesium Lanthanide Borates \ce{Mg\textit{Ln}B5O10}}
    \end{center}
}

\renewcommand{\thefigure}{S\arabic{figure}}
\setcounter{figure}{0}
\renewcommand{\thetable}{S\arabic{table}}
\setcounter{table}{0}
\renewcommand{\thesection}{\Roman{section}}
\setcounter{section}{0}
\renewcommand{\thepage}{\roman{page}}
\setcounter{page}{1}

\section{Reagents Used}

\begin{table}[htbp]
    \centering
    \begin{tabular}{ccc}
    \hline
    \textbf{Reagent}&\textbf{Supplier}&\textbf{$\%$ Purity} \\
    \hline
    \ce{La2O3}&ACROS Organics&99.99 \\
    \ce{Pr6O11}&Alfa Aesar REacton&99.996 \\
    \ce{Nd2O3}&Alfa Aesar REacton&99.99 \\
    \ce{Sm2O3}&REacton&99.99 \\
    \ce{Eu2O3}&Thermo Scientific&99.99 \\
    \ce{Gd2O3}&REacton&99.9 \\
    \ce{Tb2O3}&Aldrich&99.99 \\
    \ce{Dy2O3}&REacton&99.99 \\
    \ce{Ho2O3}&MSE Supplies&99.99 \\
    \ce{Er2O3}&Alfa Aesar REacton&99.9 \\
    \ce{Yb2O3}&MSE Supplies&99.999 \\
    \ce{MgO}&Alfa Aesar Puratronic&99.995 \\
    \ce{H3BO3}&Alfa Aesar&99.99 \\
    PVA&Sigma& Unknown \\
    \ce{HNO3}&Fisher Scientific&70\%\ aq. soln. \\
    \ce{Mg(CH3COO)2}$\cdot$\ce{4H2O}&Alfa Aesar&98 \\
    \ce{C6H8O7}&Alfa Aesar&99.5$+$ \\
    \ce{Ho(NO3)3}$\cdot$\ce{5H2O}&Aldrich&99.9 \\
    \ce{Ag} powder&Alfa Aesar&99.99, --635 mesh \\
    \hline
    \end{tabular}
    \caption{Reagents used for this investigation.}
    \label{tab:reagents}
\end{table}

\section{Solid-state Synthesis Attempts}

Prior to adopting a sol-gel synthesis as reported by \cite{zhang_phase_2017}, solid-state synthesis was also attempted. Initial attempts to produce \ce{Mg\textit{Ln}B5O10} followed the method of Cascales \textit{et al.} \cite{cascales_paramagnetic_1999} Solid \textit{Ln} oxide (\ce{\textit{Ln}2O3}), \ce{MgO} and \ce{H3BO3} were ground in a mortar and heated in an alumina crucible in air to 700--1000~$^{\circ}$C. The percentage excess of \ce{MgO} and/or \ce{H3BO3}, the furnace temperature, heating rate and reaction time were all varied and the solid-state method was attempted for \textit{Ln} = La, Nd, Pr, Gd, Tb, Ho and Yb. 

Following Mukherjee \textit{et al.} \cite{mukherjee_magnetic_2018,mukherjee_magnetic_2017}, a 2 hour 350~$^{\circ}$C preheat step was also attempted. This step decomposes the \ce{H3BO3} to \ce{B2O3}, before re-grinding and performing the high temperature reaction. Solid-state synthesis attempts ultimately failed to produce \ce{Mg\textit{Ln}B5O10} samples of sufficient purity. 

For \ce{H3BO3}, \ce{MgO} excesses of 5$\%$, the 350~$^{\circ}$C preheat step marginally improved product yield. However, the same yield or better could be achieved at 1000~$^{\circ}$C with 10$\%$ excess, with or without the preheat. 

Higher synthesis temperatures were not attempted, since \ce{Mg\textit{Ln}B5O10} melting points fall between $\sim$1000--1100~$^{\circ}$C. \cite{saubat_synthesis_1980} At lower temperatures, \ce{\textit{Ln}(BO2)3} was suppressed, but the \ce{\textit{Ln}BO3} ($C2/c$, $P63/m$) phases emerged instead. For \textit{Ln} = Gd, a triclinic $\nu$\ce{GdBO3} (P-1) phase (with lattice parameters similar to those reported by Meyer \cite{meyer_trikline_1972}) became the dominant impurity. $\nu$\ce{GdBO3} is stable at high pressures, but evidently metastable at 800--900~$^{\circ}$C after 24 hours. \cite{meyer_trikline_1972}

An increased heating rate (from 3~$^{\circ}$C/min to 10~$^{\circ}$C/min) was tried at the most promising temperature of 1000~$^{\circ}$C, in the hope that impurity phases would have insufficient time to form. (Impurity percentages generally increased for prolonged heating times.) This gave a negligible improvement compared to previous 1000~$^{\circ}$C samples.

\begin{figure}[htbp]
    \centering
    \includegraphics[width=0.8\linewidth]{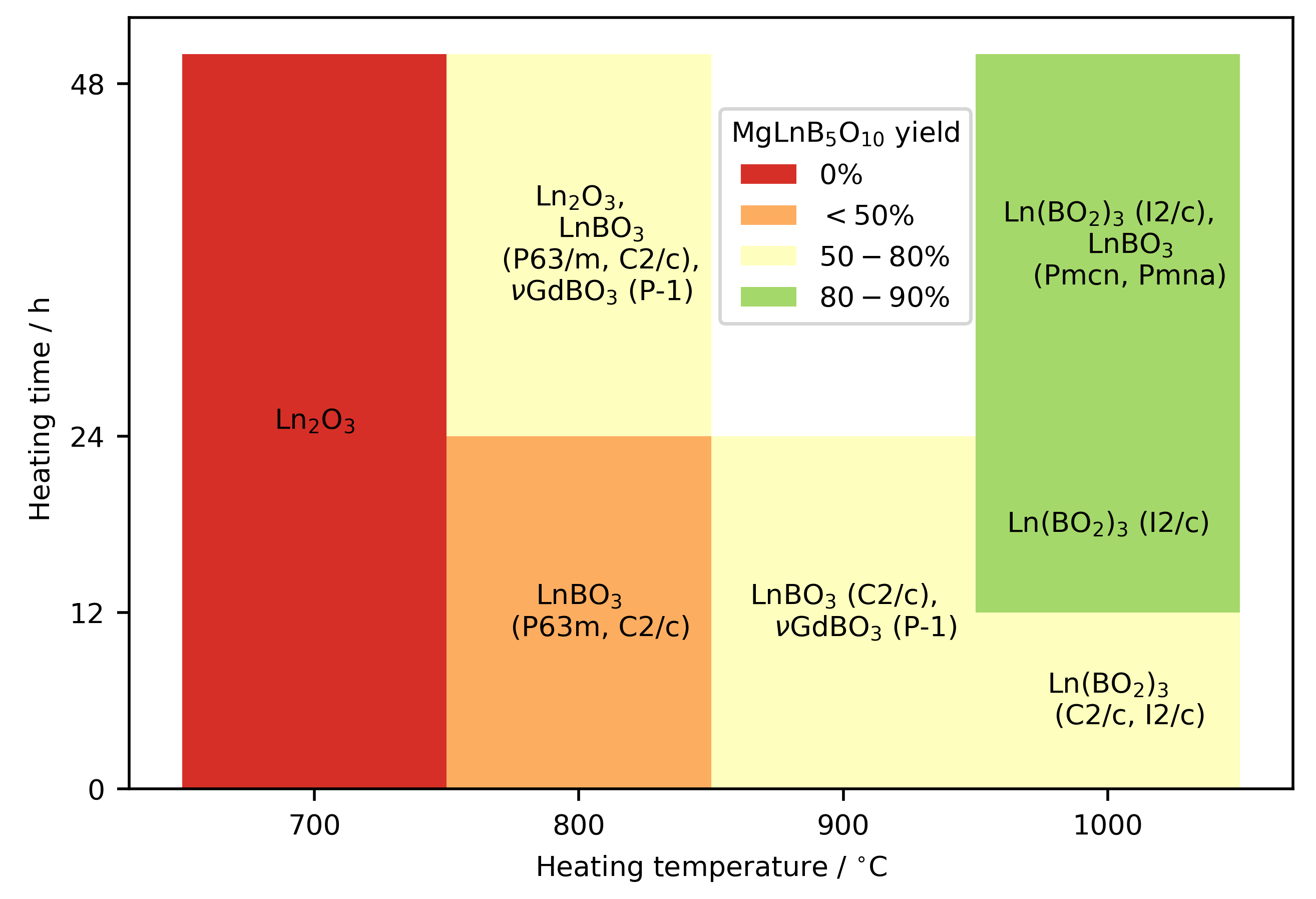}
    \caption{\ce{MgLaB5O10} yield (red-green) under different heating profiles. The dominant impurities for each heating regime are listed. \ce{Ln2O3} only completely reacted at/above 800~$^{\circ}$C. Lower temperature samples contained \ce{\textit{Ln}BO3} as the dominant impurity, with \ce{\textit{Ln}(BO2)3} being favoured at higher temperature.}
    \label{fig:solid_state}
\end{figure}

A graphical illustration of the product yield and impurity phases is given in figure~\ref{fig:solid_state}. Synthesis attempts began with \ce{Mg\textit{Ln}B5O10} at 1000~$^{\circ}$C, following Saubat \textit{et al.}, \cite{saubat_synthesis_1980} and $\sim$90$\%$ yield was obtained after 24 hours. Further heating suppressed the \ce{\textit{Ln}(BO2)3} impurity, but \ce{\textit{Ln}BO3} ($Pmcn/Pmna$) then began to form. Increasing the stoichiometric excess of \ce{H3BO3} and \ce{MgO} from 5\%\ to 10\%\ had a negligible effect on product yield, but a further increase to 20\%\ decreased the yield to $\sim 75$\%\ for some \textit{Ln}.

\section{Fitting to the Brillouin Function}
In the Weiss model, the argument of the Brillouin function is $y = g_J \mu_B J (B+\lambda M)/k_B T$, to first order in $\lambda$, with positive and negative $\lambda$ indicating ferro- and antiferromagnetism respectively. \cite{blundell_magnetism_2001} Since the argument of the Brillouin function contains the magnetisation, the system would have to be solved iteratively/self-consistently. (An example of this approach can be found in \cite{laouyenne_magnetic_2016}.)

Alternative approaches to fitting data to the Brillouin function with spin interactions include modifying $y \to g_J \mu_B J B/k_B (T+T_0)$ \cite{anderson_magnetization_1986} and allowing either the spin $J$ or the $g$ factor to vary as a fitting parameter (\cite{butler_magnetic_2021} and \cite{davis_tunable_2024} respectively). 

In this investigation, fits were attempted with the $g$ factor varying, in the hope that this would fix the curvature and provide accurate fitted values of the saturation magnetisation $M_s$. The presence of ferro/antiferromagnetic interactions could then be inferred from an increase/decrease in the fitted $g$ factor from $g_J$. However, the fit qualities proved poor. It is possible that with three variable parameters ($M_s$, $g$, $\chi_1$), the fitting algorithm was converging on a local minimum in least-squares space, rather than the global minimum.

Free spins fits were also attempted ($g = g_J$). Ideally, given the true $M_s$, one could draw qualitative conclusions on ferro- or antiferromagnetic exchange interactions from the data deviating above/below the modified free spins Brillouin function (equation 5 in main text). However, with $M_s$ unknown, the algorithm fits the data on both sides of the modified free spins Brillouin function to obtain a better fit. Hence, conclusions about any exchange interactions cannot be drawn, and the fitted $M_s$ value is inaccurate. 

\begin{figure}[htbp]
    \centering
    \includegraphics[width=0.5\linewidth]{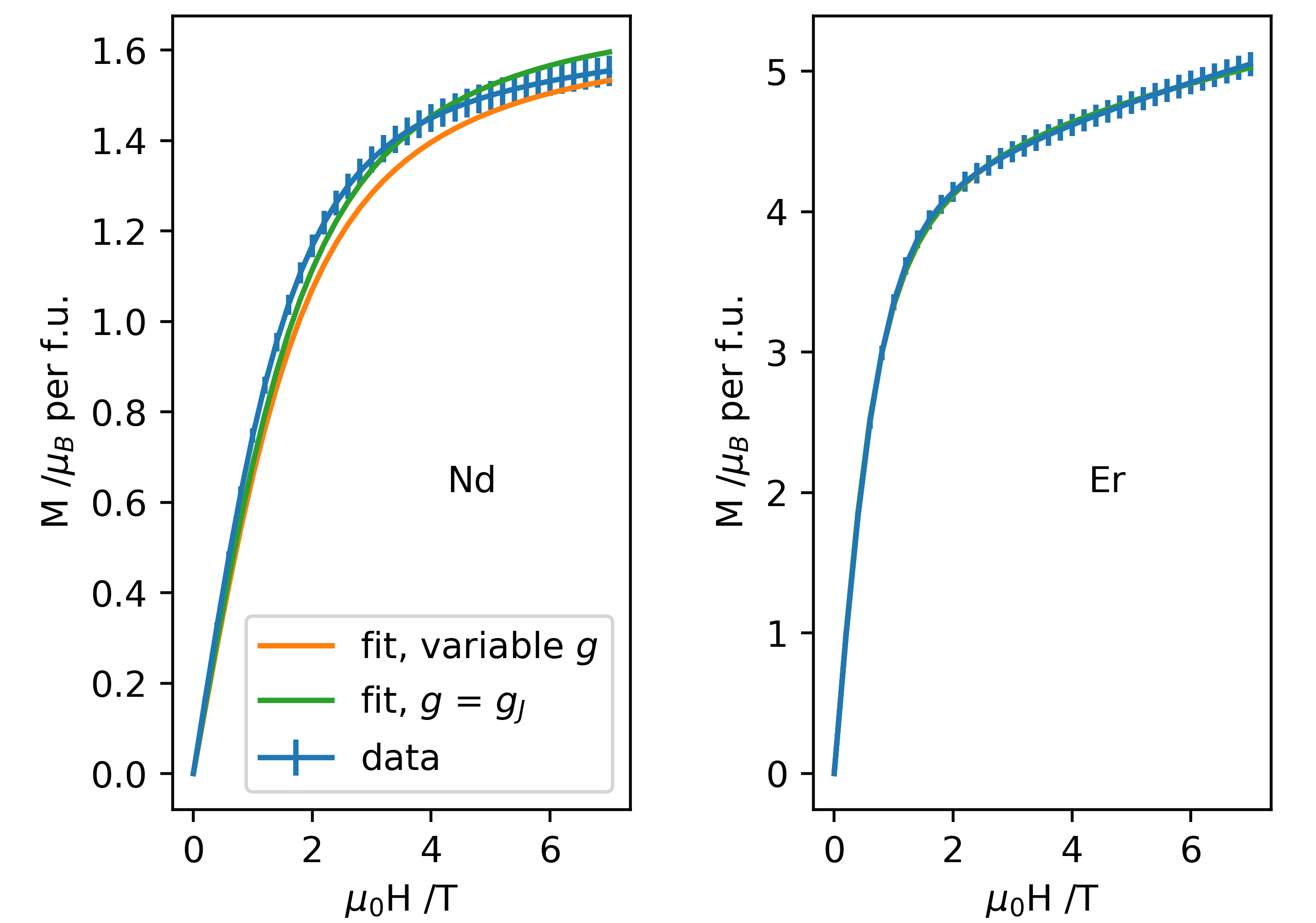}
    \caption{Magnetisation $M(H)/\mu_B$ per formula unit at 2 K, with fits to the modified Brillouin function assuming free spins ($g = g_J$) and with variable $g$ factor. Fit quality with variable $g$ was generally poor. An improved fit was obtained with the free spins Brillouin function, but at the expense of inaccuracy in determining the saturation magnetisation $M_s$. An exception is \ce{MgErB5O10}, with a good fit to the free spins modified Brillouin function. (All three curves coincide.)}
    \label{fig:Brillouin}
\end{figure}

The above is illustrated with \ce{MgNdB5O10} in figure~\ref{fig:Brillouin}. By contrast, the isothermal magnetisation data for \ce{MgErB5O10} fit the free spins Brillouin function well - so it can be inferred that the other \textit{Ln} in Table 3 (main text) do have some form of spin interaction.

For many \textit{Ln}, the fitted $\chi_1$ values were arbitrarily small but the magnetisation was still increasing, implying a still rising slope of the Brillouin function and hence incomplete spin saturation at 7 T. However, this might simply be an artefact of the poor fitting. 

The data in Table 3 (main text) represent the spread of $M_s$ and $\chi_1$ values obtained from both fitting methods. The $M_s$ estimates were sufficient to distinguish between Heisenberg and Ising spin systems. For all \textit{Ln} except \ce{Er}, the error in $\chi_1$ was comparable to $\chi_1$ itself, so conclusions about the extent of van Vleck paramagnetism in those systems cannot be made.

\end{document}